# Constraints on $\Omega$ from the *IRAS* Redshift Surveys

Shaun Cole[1,3], Karl B. Fisher[2,4] and David H. Weinberg[2,5]
[1] *Department of Physics, University of Durham, Science Laboratories, South Rd, Durham DH1 3LE*
[2] *Institute for Advanced Study, Olden Lane, Princeton, NJ 08540, USA*
[3] *Shaun.Cole@durham.ac.uk*
[4] *Fisher@guinness.ias.edu*
[5] *dhw@guinness.ias.edu*

**ABSTRACT**
We measure the anisotropy of the redshift-space power spectrum in the 1.2-Jy and QDOT redshift surveys of *IRAS*-selected galaxies. On large scales, this anisotropy is caused by coherent peculiar motions, and gravitational instability theory predicts a distortion of the power spectrum that depends only on the ratio $\beta \equiv f(\Omega)/b \approx \Omega^{0.6}/b$, where $\Omega$ is the cosmological density parameter and $b$ is the bias parameter. On small scales, the distortion is dominated by the random velocity dispersion in non-linear structures. We fit the observed anisotropy with an analytic model that incorporates two parameters, $\beta$, and a small-scale velocity dispersion $\sigma_v$. Tests on N-body simulations show that this model recovers $\beta$ quite accurately on the scales accessible to the existing *IRAS* redshift surveys. Applying our procedure to the 1.2-Jy and QDOT surveys, we find $\beta = 0.52 \pm 0.13$ and $\beta = 0.54 \pm 0.3$, respectively. These results imply $\Omega \approx 0.35$ if galaxies trace mass, or a bias factor of about 2 if $\Omega = 1$.

## 1 INTRODUCTION

If galaxies were expanding in perfect Hubble flow, then a galaxy's redshift would give a precise indication of its distance, and the clustering observed in galaxy redshift surveys would be statistically isotropic. But the Hubble flow is not perfect; peculiar velocities shift the apparent distances of galaxies, making the line of sight a preferred direction in redshift space. On small scales, velocity dispersions in collapsing and virialized systems create "fingers of God" that smear out structure along the line of sight. On large scales an opposite effect occurs; coherent flows into high-density regions and out from low-density regions enhance structures along the line of sight. Clustering anisotropies encode information about the density parameter $\Omega$, which relates gravitational forces to mass fluctuations $\delta\rho/\rho$, and about "bias" of the galaxy distribution, which *isotropically* amplifies clustering of galaxies relative to mass. In this paper, an extension of our earlier work (Cole, Fisher & Weinberg 1993, hereafter Paper I), we measure the anisotropy of the redshift-space power spectrum in the 1.2-Jy (Fisher et. al. 1995) and QDOT (Lawrence et. al. 1994) surveys of *IRAS* galaxies, and we derive corresponding constraints on $\Omega$ and the biasing parameter $b$.

Early papers on the interpretation of redshift-space anisotropy include discussions of the "cosmic virial theorem" (Peebles 1975), models for the two-point correlation function in redshift space (Rivolo & Yahil 1981; Davis & Peebles 1983; Bean et. al. 1983), and the discussion of large-scale amplification by Sargent & Turner (1977). Most recent work on the subject starts either from Peebles' (1980, § 76) analysis of the redshift-space correlation function or from Kaiser's (1987) delightfully simple formula for the redshift-space power spectrum,

$$P^S(k,\mu) = P^R(k)(1+\beta\mu^2)^2, \quad \beta \equiv f(\Omega)/b \approx \Omega^{0.6}/b. \quad (1.1)$$

Here $P^R(k)$ is the real-space power spectrum, and $P^S(k,\mu)$ is the redshift-space power spectrum, with $k$ the wavenumber and $\mu$ the cosine of the angle between the wavevector and the line of sight. Equation (1.1) assumes linear perturbation theory and a linear bias between galaxies and mass,

$$\frac{\delta N}{N} = b\frac{\delta M}{M}, \quad (1.2)$$

both of which are likely to be adequate approximations for the power spectrum on sufficiently large scales. Equation (1.1) also assumes that the observer is far away from the structure being measured, since it is only in this limit that an angle between a wavevector and the line of sight is well defined (Paper I; Zaroubi & Hoffman 1995). The relation between the Peebles and Kaiser formulations of the redshift-space anisotropy is discussed in Fisher (1995).

In Paper I we proposed a practical way to apply Kaiser's analysis to galaxy redshift surveys, by measuring $P^S(k,\mu)$ in spherical subsamples that subtend a small solid angle on the sky. Even for opening angles as large as $50°$, the (calculable) geometrical corrections to the distant-observer approximation are smaller than 5%. At each wavenumber $k$, the angular dependence of the power spectrum can be expressed in terms of a multipole decomposition,

$$\begin{aligned} P^s(k,\mu) &= \sum_{l=0}^{\infty} P_l(k) L_l(\mu) \\ &= P_0(k) + \frac{1}{2}\left(3\mu^2 - 1\right) P_2(k) + \ldots, \end{aligned} \quad (1.3)$$

where $L_l(\mu)$ is the $l^{\rm th}$ Legendre polynomial. Orthogonality of the Legendre polynomials leads to the relation

$$P_l(k) = \frac{2l+1}{2} \int_{-1}^{+1} d\mu \, P^s(k,\mu) L_l(\mu) \quad . \quad (1.4)$$

Symmetry about the line of sight implies that the odd multipoles must vanish in any statistically fair sample, and for the linear-theory distortion of equation (1.1) only the monopole, quadrupole, and hexadecapole moments are non-zero. In the linear regime, the quadrupole-to-monopole ratio $P_2(k)/P_0(k)$ provides a simple estimator for $\beta$ that is independent of the underlying real-space power spectrum. [One can also construct linear-theory estimators involving the hexadecapole $P_4(k)$, but in practice we find that measurements of $P_4(k)$ are too noisy to be useful.] In Paper I we described how the harmonic moments $P_2(k)$ and $P_0(k)$ may be estimated from windowed redshift data, with corrections for the systematic effects of shot noise and Fourier-space convolution. We also examined the behaviour of $P^s(k,\mu)$ in cosmological N-body simulations, in order to study departures from linear theory.

Paper I concluded with an application of our technique to the 1.2-Jy *IRAS* survey. However, we described this application as preliminary for three reasons. First, we had no estimated error on our derived value of $\beta$. Second, at each wavelength we used only those data in the closest shell for which our spherical window satisfied the small-angle constraint. Third, we used a linear-theory estimator to derive $\beta$ from the moments of $P^s(k,\mu)$, but our N-body tests showed that the non-linear effects of small-scale velocity dispersions were important on all scales that we could reliably probe with the 1.2-Jy survey.

In this paper we address all three of these points, developing an extended formalism and applying it to the 1.2-Jy and QDOT *IRAS* surveys. We deal with velocity dispersion by fitting a two-component model that combines linear-theory amplification of the power spectrum with suppression from a random, small-scale velocity dispersion. This approach is similar in spirit to that adopted in Fisher et al.'s (1994b) analysis of the redshift-space correlation function: we remove the small-scale dispersion effects by model fitting and use the residual large-scale distortion to constrain $\beta$. We adopt the formalism of Feldman, Kaiser & Peacock (1993; hereafter FKP) to derive the statistical error in the windowed estimates of $P^s(k,\mu)$ as a function of the window distance. These error values allow us to combine all of our data in a sensible way to obtain global estimates of $P^s(k,\mu)$, instead of using a single shell of data at each $k$ as we did in Paper I. Finally, we estimate errors in our derived value of $\beta$ using mock catalogues drawn from N-body simulations.

Section 2 describes our 2-parameter model for the multipole moments of $P^s(k,\mu)$. We use the Zel'dovich (1970) approximation and N-body simulations to test the extent to which this model provides an adequate description of power spectrum anisotropy. Section 3 describes our statistical methodology. We derive the appropriate statistical weights for different windows following FKP, and we describe how we estimate $\beta$ and the small-scale dispersion from individual estimates of $P_2(k)$ and $P_0(k)$. In § 4 we apply these techniques to mock catalogues in order to test for systematic biases and estimate the uncertainties expected from surveys the size of the 1.2-Jy and QDOT *IRAS* samples. We also apply the method to much larger mock catalogues ($\sim 10^6$ galaxies) in order to see the reduction of statistical errors that can be expected in future redshift samples like the Sloan Digital Sky Survey (Gunn & Knapp 1993; Gunn & Weinberg 1995) and the Anglo-Australian Telescope 2dF galaxy survey. In § 5 we apply the technique to the actual 1.2-Jy and QDOT redshift surveys. We discuss the prospects for this power spectrum technique and other approaches to clustering anisotropy in § 6.

## 2 A 2-PARAMETER MODEL FOR REDSHIFT-SPACE ANISOTROPY

In Paper I we studied the non-linear effects that cause the redshift-space distortion measured in N-body simulations to deviate from the prediction of linear theory. Although there are several distinct sources of non-linear behaviour, the most troublesome is the "dispersion non-linearity" caused by random galaxy velocities in clusters and other collapsed structures, because this has an important influence on all of the scales that can be probed accurately with current redshift surveys. Below we present a simple, 2-parameter, analytic model of the effect of random velocities on the redshift-space power spectrum. We assess its accuracy by comparison with large volume simulations carried out using the Zel'dovich approximation and N-body simulations of somewhat smaller volumes.

### 2.1 Analytic Model

The model that we consider is one in which galaxy velocities are given by linear theory plus an uncorrelated, random velocity dispersion. In reality the non-linear velocity field will be correlated with the density field, but our simple model may be a useful first approximation. Peacock & Dodds (1994) discussed a model of this form and showed that Gaussian distributed random velocities added to the linear-theory flow transform the redshift-space power spectrum to

$$P^s(k,\mu) = P^R(k)\,(1+\beta\mu^2)^2\,\exp(-k^2\sigma_v^2\mu^2). \quad (2.1)$$

In general, the velocity distribution of galaxies identified in N-body simulations is more nearly exponential than Gaussian, and an exponential distribution also offers a better fit to the observed redshift-space correlation function (Davis & Peebles 1983; Fisher et. al. 1994b). If one adopts an exponential distribution for the random velocities, then the corresponding expression for the power spectrum is (Park et. al. 1994)

$$P^S(k,\mu) = P^R(k)\,(1+\beta\mu^2)^2\,(1+k^2\sigma_v^2\mu^2/2)^{-2}. \tag{2.2}$$

In each case one recovers equation (1.1) in the limit of $k\sigma_v \to 0$. Beyond this the two expressions look quite different, but in fact they differ only at order $\mathcal{O}(k^4\sigma_v^4)$. This similarity is reassuring, for it means that at long wavelengths the effect of random velocities on the redshift-space power spectrum depends primarily on the second moment of the velocity distribution $\sigma_v$ and is insensitive to the form of the velocity distribution function.

It is tedious but straightforward to compute the monopole and quadrupole components of the redshift-space power spectra resulting from the Gaussian and exponential models. The ratio of the redshift-space monopole to the underlying real-space power spectrum can be written in the form

$$\frac{P_0(k)}{P^R(k)} = A(\kappa) + \frac{2}{3}\beta\,B(\kappa) + \frac{1}{5}\beta^2\,C(\kappa) \quad, \tag{2.3}$$

where $\kappa \equiv k\sigma_v$. In the Gaussian model

$$\begin{aligned}
A(\kappa) &= \frac{\sqrt{\pi}}{2}\frac{\mathrm{Erf}(\kappa)}{\kappa}, \\
B(\kappa) &= \frac{3}{2\kappa^2}\left(A(\kappa) - e^{-\kappa^2}\right), \\
C(\kappa) &= \frac{5}{2\kappa^2}\left(B(\kappa) - e^{-\kappa^2}\right),
\end{aligned} \tag{2.4}$$

while in the exponential model

$$\begin{aligned}
A(\kappa) &= \frac{\tan^{-1}(\kappa/\sqrt{2})}{\sqrt{2}\kappa} + \frac{1}{2+\kappa^2}, \\
B(\kappa) &= \frac{6}{\kappa^2}\left(A(\kappa) - \frac{2}{2+\kappa^2}\right), \\
C(\kappa) &= -\frac{10}{\kappa^2}\left(B(\kappa) - \frac{2}{2+\kappa^2}\right) \quad.
\end{aligned} \tag{2.5}$$

The expression for the quadrupole moment in the Gaussian model is

$$\begin{aligned}
\frac{P_2(k)}{P^R(k)} &= \frac{5}{2}[B(\kappa) - A(\kappa)] + \beta\left[\frac{4}{3}B(\kappa) + 3(C(\kappa) - B(\kappa))\right] \\
&\quad + \beta^2\left[C(\kappa) + \frac{15}{4\kappa^2}(C(\kappa) - B(\kappa))\right],
\end{aligned} \tag{2.6}$$

with $A$, $B$, and $C$ given by equations (2.4). The quadrupole moment in the exponential model is

$$\begin{aligned}
\frac{P_2(k)}{P^R(k)} &= \frac{5}{2}[B(\kappa) - A(\kappa)] + \beta\left[\frac{4}{3}B(\kappa) + 3(C(\kappa) - B(\kappa))\right] \\
&\quad + \beta^2\left[\frac{5}{\kappa^2}(B(\kappa) - C(\kappa))\right],
\end{aligned} \tag{2.7}$$

with $A$, $B$, and $C$ given by equations (2.5).

For $\kappa \ll 1$, both equations (2.4) and (2.5) have series expansions given by

$$\begin{aligned}
A(\kappa) &\approx 1 - \frac{1}{3}\kappa^2 \\
B(\kappa) &\approx 1 - \frac{3}{5}\kappa^2 \\
C(\kappa) &\approx 1 - \frac{5}{7}\kappa^2 \quad.
\end{aligned} \tag{2.8}$$

As $\kappa \to 0$, $A$, $B$, and $C$ go to unity, and equations (2.3), (2.6), and (2.7) yield the familiar linear theory results for the monopole and quadrupole (cf. Paper I, equation 2.3).

### 2.2 Comparison with the Zel'dovich Approximation

We now compare the analytic description of the redshift-space power spectrum with the results of a set of simulations constructed by combining the Zel'dovich approximation with a random small-scale velocity dispersion. We will turn to tests on N-body simulations shortly, but there are two reasons to begin with the Zel'dovich models. First, since the Zel'dovich approximation reduces to linear theory in the limit of small fluctuation amplitude, we can construct numerical realizations that should correspond precisely to our analytic model. Second, with the Zel'dovich approximation we can do enough large-volume simulations to obtain small statistical errors on the large scales where linear theory begins to hold. The Zel'dovich approximation is valid in the linear and mildly non-linear regimes, so while we cannot trust the non-linear behaviour in detail, we can get an idea of the scales at which non-linear effects other than the small-scale velocity dispersion (which is in the models by construction) are likely to become significant.

We start the Zel'dovich simulations from Gaussian initial conditions constructed on a $200^3$ periodic grid, with a power spectrum given by the formula of Efstathiou, Bond & White (1992) with their scale parameter set to $\Gamma = 0.25$. The shape of this power spectrum is consistent with recent observations of large-scale galaxy clustering and in particular with the clustering of *IRAS* galaxies (Efstathiou et. al. 1990; Saunders et. al. 1991; Fisher et. al. 1993; FKP; Moore et. al. 1994). In order to span a wide range of scales, we perform simulations with three different box sizes, $l_{\mathrm{box}} = 1200, 400$ and $200h^{-1}\mathrm{Mpc}$. Truncation of the initial power spectrum at the Nyquist frequency of the grid leads to spurious small-scale effects in the large boxes, so we need the smaller boxes in order to probe the scales where the velocity dispersion effects become important. The largest boxes give us good sampling of the long wavelength modes that are near the linear regime.

The Zel'dovich approximation yields positions and velocities of the $200^3$ particles in each simulation. We add to each particle's velocity a random component drawn from a Gaussian distribution (in each of the three spatial dimensions) with 1-d dispersion of $\sigma_v = 300\,\mathrm{km\,s}^{-1}$. We compute the redshift-space density field as it would appear to an observer at $x = -\infty$ by cloud-in-cell weighting the particle distribution onto a $128^3$ grid, defining each particle's redshift-space position to be $(x+v_x/H_0, y, z)$, where $v_x$ is the $x$-component of peculiar velocity. This procedure preserves the periodic boundary condition, and we use an FFT to obtain the Fourier modes of the density field on a Cartesian grid. If the clustering were isotropic, then the power spectrum would depend only on the modulus of $k$, but redshift-space distortions introduce anisotropy. At each value of $k$, we decompose the power spectrum into multipole moments (equation 1.3), extracting the multipole coefficients $P_l(k)$ by least-squares fitting. We repeat the procedure for the $y$ and $z$ coordinate axes, then average the multipoles obtained from the three viewing directions to reduce the statistical noise.

Figure 1 displays the results of this analysis. Open symbols show the quadrupole-to-monopole ratio from simulations with $\Omega = 1$ (circles) and $\Omega = 0.3$ (squares), with the linear-theory power spectrum normalized so that the *rms*

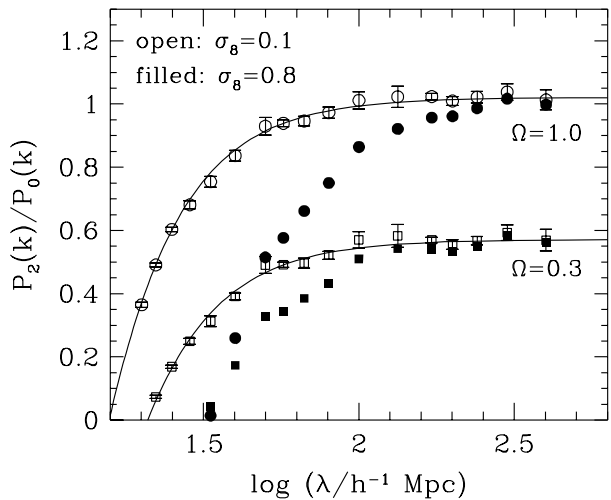

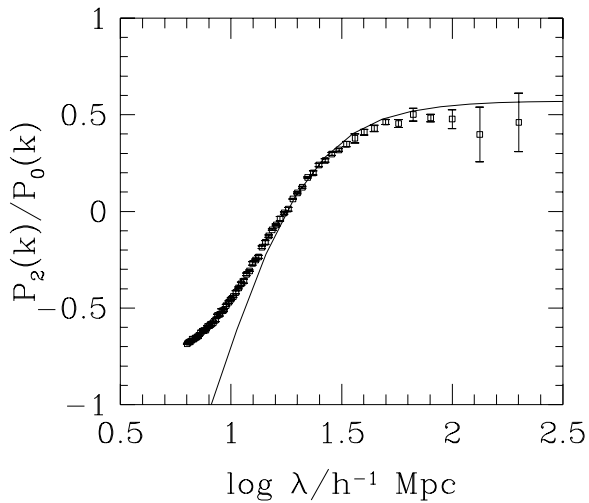

**Figure 1.** The quadrupole-to-monopole ratio, $P_2(k)/P_0(k)$, of the redshift-space power spectrum as a function of wavelength. The symbols and error bars show the mean and error on the mean of estimates of this ratio made from the Zel'dovich approximation simulations described in § 2.2. Points with $\log \lambda > 2.2$ come from the $1200 h^{-1}$ Mpc boxes, points with $1.7 < \log \lambda < 2.2$ from the $400 h^{-1}$ Mpc boxes, and points with $\log \lambda < 1.7$ from the $200 h^{-1}$ Mpc boxes. Open symbols represent a low fluctuation amplitude, $\sigma_8 = 0.1$, and filled symbols a fluctuation amplitude $\sigma_8 = 0.8$. Circles are for $\Omega = 1$ and squares for $\Omega = 0.3$. Solid curves show the analytic model with $\beta = \Omega^{0.6}$ ($b = 1$ in all cases) and $\sigma_v = 300 \, \mathrm{km\,s^{-1}}$, the value of the small-scale dispersion used in the simulations. For $\sigma_8 = 0.1$ the analytic model fits the numerical results perfectly, but for $\sigma_8 = 0.8$ non-linear dynamical evolution influences the results out to quite large scales, roughly a 10% effect at $\lambda = 100 h^{-1}$ Mpc.

**Figure 2.** The quadrupole-to-monopole ratio, $P_2(k)/P_0(k)$, as a function of wavelength. The symbols and error bars show estimates of the mean and error on the mean of measurements made from an ensemble of four N-body simulations. The simulations model a periodic box of $l_{\mathrm{box}} = 400 \, h^{-1}$ Mpc with $\Omega = 0.3$ and $\sigma_8 = 0.8$. The solid curve is the exponential model of § 2.1 for $\beta = \Omega^{0.6} = 0.486$ and $\sigma_v = 280 \, \mathrm{km\,s^{-1}}$

mass fluctuation in spheres of radius $8 h^{-1}$ Mpc is $\sigma_8 = 0.1$. For each model and box size we average the results of eight independent simulations; error bars indicate the $1 - \sigma$ uncertainty in the mean value, i.e. the run-to-run dispersion divided by $\sqrt{N-1} = \sqrt{7}$. Solid curves show the analytic prediction for the Gaussian velocity distribution, obtained from equations (2.3) and (2.6). Since the fluctuation amplitude in these runs is low, we expect good agreement between the numerical and analytic results, and indeed the agreement in Figure 1 is essentially perfect.

Filled symbols in Figure 1 show the quadrupole-to-monopole ratio for Zel'dovich simulations that are identical except for a higher normalization of the linear-theory power spectrum, $\sigma_8 = 0.8$. This normalization is intermediate between the value determined from optical galaxy surveys, $\sigma_8 \sim 1$ (Davis & Peebles 1983), and the value from the *IRAS* surveys, $\sigma_8 \sim 0.7$ (Fisher et. al. 1994a). Differences between the solid and filled symbols arise from non-linear evolution. These non-linear effects – which are beyond those caused by our random, small-scale dispersion – are present at the $10 - 15\%$ level for $\lambda \approx 100 h^{-1}$ Mpc, and they can be detected out to wavelengths as large as $200 h^{-1}$ Mpc. To some extent, the additional non-linearities have the same effect as raising the small-scale velocity dispersion, which is what we might expect after shell-crossing in the Zel'dovich approximation.

### 2.3 Comparison with N-body Simulations

The Zel'dovich realizations provide good statistics on large scales, so we can get an idea of where non-linearities other than the velocity dispersion effect will become important. However, after structure becomes non-linear, we cannot assume that the Zel'dovich approximation offers a reliable description. To explore non-linear departures from our 2-parameter model more completely, we turn to N-body simulations.

The simulations have Gaussian initial conditions, again with the $\Gamma = 0.25$ power spectrum of Efstathiou et. al. (1992). Each of the four independent simulations models a periodic box of size $l_{\mathrm{box}} = 400 \, h^{-1}$ Mpc. We use a staggered-mesh PM code written by C. Park (1990) to evolve a density field represented by $200^3$ particles, with a $400^3$ mesh for force computations. We adopt an open cosmology, with $\Omega = 0.3$ at $z = 0$, and we normalize the initial power spectrum so that linear-theory evolution to $z = 0$ would yield an *rms* fluctuation in $8 h^{-1}$ Mpc spheres of $\sigma_8 = 0.8$. We begin the simulations at a redshift of 24 and evolve to the present in 48 equal steps, using the expansion factor $a$ as the time variable for integration. The large timesteps are adequate because of the rather low ($\sim 1 - 2 h^{-1}$ Mpc) force resolution of the computations; our goal is to model the large-scale density and velocity fields, not the internal structure of collapsed objects.

We analyse the redshift-space power spectra of the simulations in the same manner as described for the Zel'dovich simulations. The resulting quadrupole-to-monopole ratios are shown in Figure 2 along with a curve showing the exponential model of § 2.1. The curve is for the true value of $\beta = \Omega^{0.6} = 0.486$ and a value of $\sigma_v = 280 \, \mathrm{km\,s^{-1}}$, chosen by eye to give a reasonable fit to the N-body data. The Gaus-sian model with the same parameters is virtually identical to the exponential model for $\lambda > 10\,h^{-1}\,\mathrm{Mpc}$.

Because our model ignores the correlation of the velocity dispersion with the density field, there is no simple relation between the value of $\sigma_v$ deduced from the fit and velocity dispersions measured directly from the N-body simulation. In the idealized limit that the mean streaming motions of galaxies vanish and the dispersion is isotropic and independent of scale, then $\sigma_v$ is simply $\sigma_{pw}/\sqrt{2}$, where $\sigma_{pw}$ is the 1-d pairwise galaxy velocity dispersion. Analyses of the redshift-space correlation function (Fisher et. al. 1994b; Fisher 1995), however, indicate that actual redshift-space anisotropies are more complex, and that both the mean streaming and spatial dependence of the dispersion are non-neglible. In our N-body simulations the pairwise dispersion is $\sigma_{pw} = 370\,\mathrm{km\,s^{-1}}$ at separations of $10\,h^{-1}\,\mathrm{Mpc}$, so $\sigma_{pw}/\sqrt{2} = 262\,\mathrm{km\,s^{-1}}$ agrees reasonably well with the value $\sigma_v = 280\,\mathrm{km\,s^{-1}}$ that fits the power spectrum anisotropy in Figure 2, but there is no reason to expect precise correspondence.

Figure 2 shows quite good agreement between the analytic model and the N-body results. At large wavelengths, the N-body values of $P_2(k)/P_0(k)$ lie about 10% below the model values. By comparing to the Zel'dovich results on the same scales (Figure 1), we see that this discrepancy is caused by residual non-linearity that is not included in the 2-parameter model. We also see discrepancies between the model and the N-body results at short wavelengths, but the N-body results in this regime may be affected by finite resolution (particularly of the initial conditions), and for the purposes of this paper we are not much interested in these scales in any case. If we restrict ourselves to wavelengths $\lambda \gtrsim 15\,h^{-1}\,\mathrm{Mpc}$, it appears that the $\beta$-$\sigma_v$ model can allow us to estimate $\beta$ with systematic errors of $\sim 10\%$. We will find below that the random statistical errors for the 1.2-Jy and QDOT surveys are larger than this, so our 2-parameter description suffices for present purposes. However, for much larger future surveys such as the Sloan survey or the AAT 2dF survey, the statistical errors should be smaller than 10%; a more complete description of non-linear effects will be needed to take full advantage of these surveys. We see this as an encouraging result, since non-linear effects beyond the small-scale dispersion may provide a means of breaking the degeneracy between $\Omega$ and $b$ in the model of redshift-space distortions.

## 3 STATISTICAL METHODOLOGY

We now describe how we can use our model to estimate $\beta$ and $\sigma_v$ from redshift surveys. This estimation involves:

(i) measuring the redshift-space power spectrum $P^s(k,\mu)$ from individual subsamples of the galaxy catalogue and combining these estimates in a a way that makes statistically optimal use of the data,

(ii) extracting the monopole and quadrupole moments of $P^s(k,\mu)$, accounting properly for the effects of convolution due to the windowing of the galaxy subsamples, and

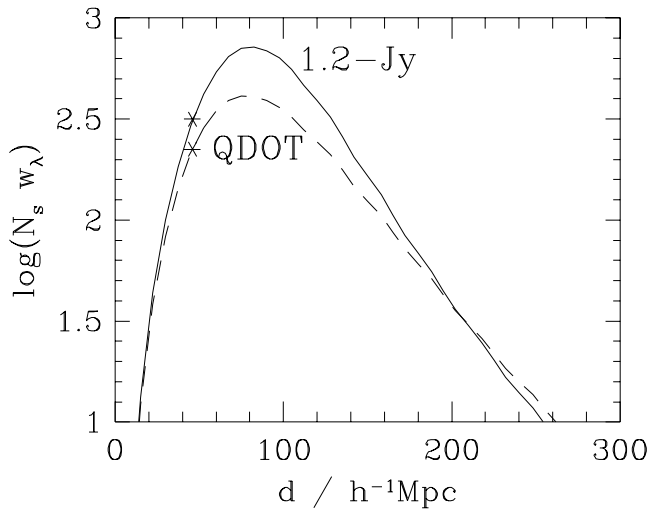

**Figure 3.** Weight functions for power spectrum estimates at the fundamental wavelength $\lambda_{\mathrm{f}} = 30\,h^{-1}\,\mathrm{Mpc}$ of bowler hat windows of radius $R_{\mathrm{sph}} = 21.5\,h^{-1}\,\mathrm{Mpc}$. Curves show the weight $w_\lambda$ given to a windowed subsample at distance, $d$, multiplied by the number of independent subsamples $N_{\mathrm{s}}$ per logarithmic interval of $d$, for the 1.2-Jy (solid line) and QDOT (dashed line) selection functions. Asterisks mark the minimum distance, $d_{\min} = 46\,h^{-1}\,\mathrm{Mpc}$, at which the windows satisfy our opening-angle constraint.

(iii) fitting the wavelength dependence of the quadrupole-to-monopole ratio to constrain the model parameters $\beta$ and $\sigma_v$.

### 3.1 Measuring $P^s(k,\mu)$

As described in Paper I, we repeatedly sample the galaxy catalogue using randomly placed spherical windows. Two constraints are used when placing the sampling window. First, the sampling window should not extend into any region not covered by the catalogue, such as the $|b| < 5°$ cut in the IRAS 1.2-Jy survey. Second, the angle subtended by the sample should be small, so that the distant-observer approximation implicit in our analytic theory of the distortions is valid. We use an apodized top-hat or "bowler-hat" window as defined in equation (3.2) of Paper I, with the ratio of the Gaussian smoothing length to top-hat radius $R_g/R_{\mathrm{sph}} = 0.1$. We choose the bowler-hat window because its close resemblance to the top-hat keeps the necessary calculations simple and the apodization reduces the sidelobes of the window in Fourier-space. The angular constraint we impose is that the diameter of the sphere subtend less than $50°$, i.e., $d_{\min} = R_{\mathrm{sph}}/\tan(25°)$. This is a compromise between making the small-angle approximation accurate and reducing the shot noise; smaller opening angles require placing the spheres at greater distances, where the sampling of the density field in flux-limited surveys is worse. As shown in Paper I, the geometrical effect of a $50°$ opening angle on the inferred quadrupole is only 5%.

We Fourier transform the galaxy distribution inside the bowler-hat window, weighting each galaxy's contribution to each Fourier mode by the inverse of the selection function, $\phi(r)$, evaluated at the galaxy's redshift.* We obtain the power spectrum $P^s(k,\mu)$, by squaring these redshift-space Fourier modes and subtracting off the white noise contribution arising from discrete sampling (shot noise), as described in § 3 of Paper I. As discussed in Paper I and Fisher et. al. (1993), it is desirable to measure the spectrum at wavenumbers that are discrete harmonics of the window function. With our bowler hat window, this restricts us to values of $k$ satisfying $j_1(kR_{sph}) = 0$, where $j_1(x)$ is the first-order spherical Bessel function. The fundamental harmonic is $k_f = 4.49/R_{sph}$ ($\lambda_f \equiv 2\pi/k_f = 1.40 R_{sph}$), and the first harmonic is $k_1 = 7.72/R_{sph}$ ($\lambda_1 = 0.81 R_{sph}$). At each value of $k$, the best estimates of $P^s(k,\mu)$ come from the closest windows that are allowed by the small-angle constraint; the noise rises significantly as the windows are displaced to greater distances because of the sparser sampling of the galaxy density field. However, there are more independent volumes at large distances. We therefore desire an optimal way to combine the estimates obtained from windows at different distances into a single estimate of $P^s(k,\mu)$. It is clear that the form of the weighting function should depend on the depth and sampling of the survey being analysed.

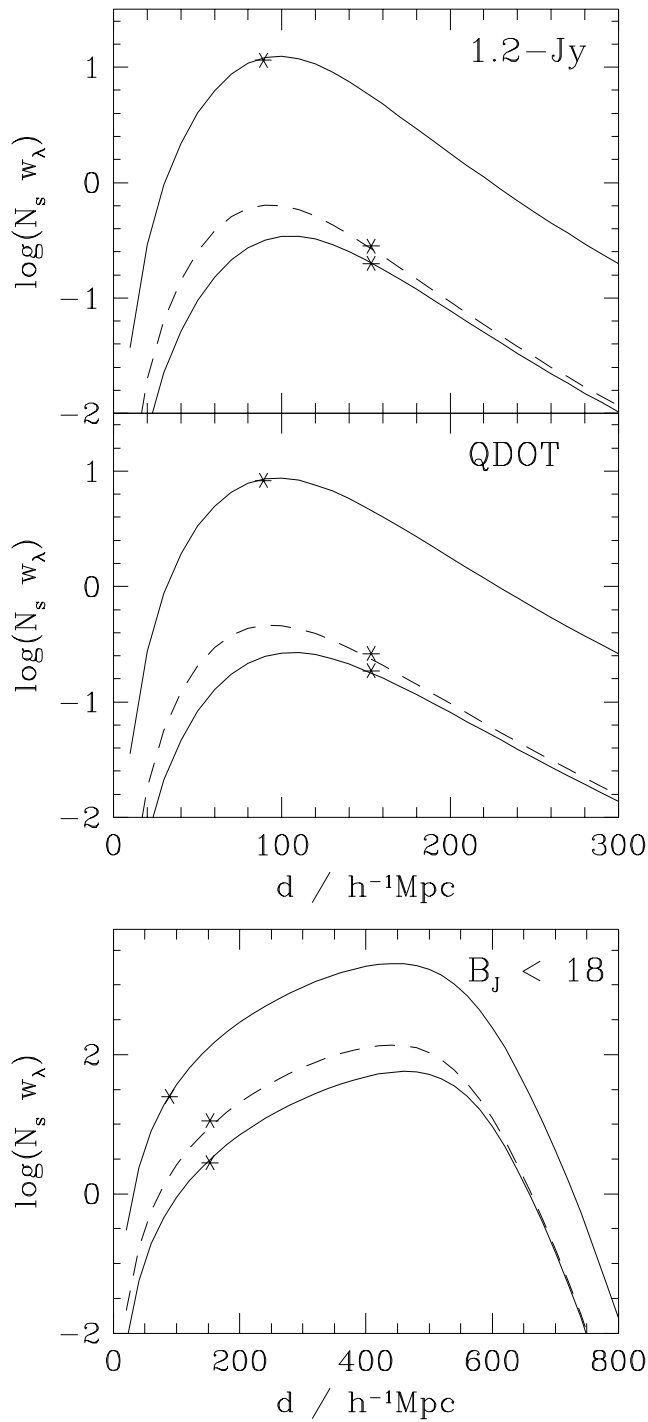

**Figure 4.** Weight functions at larger wavelengths, for selection functions of a) the 1.2-Jy survey, b) the QDOT survey, and c) a hypothetical, all-sky survey with a magnitude limit $B_J = 18$, containing just over one million galaxies. In each panel, the lower and upper solid curves represent, respectively, the fundamental wavelengths $\lambda_f = 100 h^{-1}$ Mpc and $58.65 h^{-1}$ Mpc of bowler hat windows with radii $R_{sph} = 71.5$ and $41.9 h^{-1}$ Mpc. The dashed curve is for $\lambda_1 = 58.65 \, h^{-1}$ Mpc, the first harmonic of the $R_{sph} = 71.5 h^{-1}$ Mpc sampling window. The depth that provides the largest contribution to our estimate of $P^s(k,\mu)$ increases slowly with wavelength. At fixed distance and wavelength, the estimates from the first harmonic are considerably noisier than those from the fundamental mode.

To define appropriate weights we make use of the formula for the variance, $\sigma_P^2(k)$, of the estimated power at wavenumber $k$ derived by FKP for Gaussian fluctuations. They obtain the following useful expression,

$$\sigma_P^2(k) = \frac{2}{V_k^2} \int_{\mathbf{k}} \int_{\mathbf{k'}} \Big| P(k) \, Q(\mathbf{k} - \mathbf{k'}) + S(\mathbf{k} - \mathbf{k'}) \Big|^2 \, d^3\mathbf{k} \, d^3\mathbf{k'}, \quad (3.1)$$

where the integrations are performed over a thin shell in $k$-space of volume $V_k$. The functions $Q(\mathbf{k})$ and $S(\mathbf{k})$ depend on the selection function and geometry of the galaxy sample (see FKP equations (2.4.3) and (2.4.4)). For galaxy subsamples selected according to our spherically symmetric window function $w(r)$, centred at a distance $d$, these functions are

$$Q(\mathbf{k}) = \int w^2(r) \, \exp(i\mathbf{k} \cdot \mathbf{r}) \, d^3\mathbf{r} \quad (3.2)$$

and

$$S(\mathbf{k}) = \int \frac{w^2(r)}{\bar{n}\phi(d+z)} \, \exp(i\mathbf{k} \cdot \mathbf{r}) \, d^3\mathbf{r}, \quad (3.3)$$

where $z$ is the line-of-sight component of $\mathbf{r}$, and $\bar{n}\phi(d+z)$ is the mean galaxy number density in the survey at a distance $d + z$. For convenience we have put the origin of the coordinate system at the centre of the galaxy sample. Thus $Q(\mathbf{k})$ has spherical symmetry and is independent of the window distance $d$, while $S(\mathbf{k})$ is axially symmetric and increases rapidly with distance as the galaxy number density falls in the survey. These symmetries reduce (3.1) to an integral over just three angular coordinates, greatly simplifying its evaluation.

For each choice of sampling window and wavelength, we compute $\sigma_P^2(k)$ as function of the distance $d$ of the sampling window. We then define a weight $w_\lambda(d) = 1/\sigma_P^2(k)$, which we use to combine the estimates of $P^s(k,\mu)$ determined from windows located at different distances. Figure 3 shows the weighting functions for the 1.2-Jy and QDOT selection functions at the fundamental wavelength $\lambda_f = 30 h^{-1}$ Mpc of bowler hat windows of radius $R_{sph} = 21.4 h^{-1}$ Mpc. In equation (3.1) we use the $\Gamma = 0.25$ power spectrum of Efstathiou et. al. (1992), normalized to $\sigma_8 = 0.8$. The quantity plotted is $N_s w_\lambda$, where we have defined $N_s \equiv d^3/R_{sph}^3$, which is proportional to the number of independent samples that can be taken in a logarithmic interval of distance $d$. Thus the curves of Figure 3 peak at the depth where the contribution to our overall estimate of $P^S(k,\mu)$ is largest. For both surveys this peak occurs at about $80 h^{-1}$ Mpc; the peak is higher for the 1.2-Jy survey, indicating better signal-to-noise. Asterisks mark the minimum distance that satisfies the $50°$ constraint on opening angle, $d_{min} = 46 h^{-1}$ Mpc. Since this distance lies significantly inside the peaks of the curves, eliminating the more nearby windows loses little information at this wavelength.

---

* $\phi(r)$ represents the fraction of the galaxies visible at a distance $r$ in a homogeneous flux limited survey, i.e. the number density is $\propto r^2 \phi(r)$.

Figure 4 illustrates the behaviour of the weighting functions at longer wavelengths. In each panel, the upper and lower solid lines show $N_s w_\lambda$ for the fundamental wavelengths $\lambda_f = 100$ and $58.65 h^{-1}$ Mpc of windows of radius 71.5 and $41.9 h^{-1}$ Mpc, The dashed line represents the first harmonic of the $R_{sph} = 71.5 h^{-1}$ Mpc window, which is again $\lambda_1 = 58.65 h^{-1}$ Mpc. At these wavelengths, weight functions are quite similar for the 1.2-Jy and QDOT surveys, peaking broadly near $100 h^{-1}$ Mpc. The peaks are quite a bit lower than those in Figure 3, indicating worse signal-to-noise at these larger scales. Furthermore, the positions of the asterisks at or beyond the peaks indicate that the small-angle constraint eliminates much of the potentially available information. Note also that most of the information for $\lambda = 58.65 h^{-1}$ Mpc comes from the windows where it is the fundamental wavelength rather than the first harmonic.

The bottom panel of Figure 4 shows the corresponding weight functions for a hypothetical, all-sky sample with an optical magnitude limit of $B_J = 18$. Such a sample would contain just over one million galaxies, roughly the number of redshifts expected in the Sloan survey, and about four times the number expected in the AAT 2dF redshift survey. Because of the much greater depth of this sample, the weight curves peak at distances of $500 - 600 h^{-1}$ Mpc, far outside the minimum distance imposed by the small-angle criterion, and these peaks are roughly two orders of magnitude higher than those for the 1.2-Jy and QDOT surveys. These much larger redshift surveys will yield $P^S(k,\mu)$ measurements with much higher precision, especially at wavelengths $\lambda \gtrsim 100 h^{-1}$ Mpc. The sharp decline in the weighting function at distances $> 600 h^{-1}$ Mpc is a result of the exponential cutoff in the optical luminosity function (in contrast to the power law cutoff for high luminosity *IRAS* galaxies).

### 3.2 Extracting the multipole moments $P_0(k)$ and $P_2(k)$

The weighting function allows us to combine the estimates of $P^S(k,\mu)$ obtained from windows located at various distances. We use a least-squares fit of equation (1.3) to this averaged $P^S(k,\mu)$ to extract estimates of the monopole, $P_0(k)$, and quadrupole, $P_2(k)$, moments. At this stage we must correct the monopole and quadrupole estimates for the effect of convolution. The power spectrum that we have estimated is not the the true power spectrum of the underlying galaxy population but rather the true power spectrum convolved with the square of the transform of the sampling window (cf., Paper I equation 3.4). This convolution tends both to reduce the estimated power and to make $P^S(k,\mu)$ more isotropic; if uncorrected it would lead to a systematic underestimate of the monopole and quadrupole moments. In Paper I, we quantified this effect and calculated the correction factors for the measured monopole and quadrupole. These correction factors have only a very weak dependence on the shape of the assumed real-space power spectrum. Throughout we adopt the values appropriate for the $\Gamma = 0.25$ spectrum of Efstathiou et. al. (1992), but our results are insensitive to this choice.

### 3.3 Estimation of Model Parameters

In Paper I we used the ratio $P_2(k)/P_0(k)$ to give an estimate of $\beta$ at each wavelength. Here we extend our method so that we can combine the estimates of $P_2(k)/P_0(k)$ at different wavelengths into global estimates of $\beta$ and $\sigma_v$, using the model described in § 2.1 . The great advantage of using these ratios is that the real-space power spectrum $P^R(k)$ cancels out of the model predictions, so our estimates of $\beta$ and $\sigma_v$ depend only on the observed *anisotropy* of the power spectrum and not on an assumed or estimated form of $P^R(k)$.

For each of our adopted window radii (which range from 14.3 to $100 h^{-1}$ Mpc in the *IRAS* analyses), we obtain measures of the monopole and quadrupole at the window's fundamental wavelength and first harmonic. Our data consist of the ratios $r(k) = P_2(k)/P_0(k)$ at these various wavelengths. Because of the finite size of our sampling window and the resulting convolution described above, the values of $r(k)$ at different $k$ are correlated. We can estimate $\beta$ and $\sigma_v$ by minimizing the quantity

$$\chi^2 = \sum_{i,j} \Delta_i \mathbf{M}_{ij}^{-1} \Delta_j ,$$
$$\Delta_i \equiv r_O(k_i) - r_P(k_i; \beta, \sigma_v) , \quad (3.4)$$

where $r_O(k)$ is the observed quadrupole-to-monopole ratio at wavenumber $k$, $r_P(k; \beta, \sigma_v)$ is the ratio predicted by our two-parameter model, and the covariance matrix of the data points is $\mathbf{M}_{ij} \equiv \langle (r(k_i) - \bar{r}(k_i))(r(k_j) - \bar{r}(k_j)) \rangle$, with $\bar{r}$ denoting an average over ensembles.

We estimate the covariance matrix in equation (3.4) using a series of mock catalogues drawn from N-body simulations, each tailored to the selection function and survey geometry of the catalogue under consideration. We describe the construction of these mock catalogues in § 4 below. The model we adopt for the mock catalogues affects the covariance matrix and therefore enters into the estimation of $\beta$ and $\sigma_v$, but so long as the adopted model provides a reasonably realistic representation of the data, the dependence of the final result on the details of the model should be minimal. There is one important caveat to this point, however. Quadrupole-to-monopole ratios at neighbouring values of $k$ can be quite strongly correlated, and the resulting covariance matrix can be close to singular. Since the inverse covariance matrix enters the definition of $\chi^2$, the parame-ter fits can become sensitive to small errors in the estimate of $\mathbf{M}_{ij}$. We circumvent this problem via the standard technique of principal component analysis (PCA) (e.g., Kendal 1975). Briefly, this method involves finding the linear combinations of the original data points that are statistically independent.[†] Highly correlated data points produce some combinations that are genuinely meaningful (e.g., the sum of two correlated values) and others (e.g., the difference of these values) that contain little new information but have small variances and therefore dominate the fit. In PCA, these latter combinations are discarded, and only those linear combinations that contribute significantly to the overall variance are retained. When performing our $(\beta, \sigma_v)$ fits, we first normalize all of the input data values to zero mean and unit variance, then apply PCA and keep those combinations of data points that contribute 95% of the total variance.

We should emphasize that PCA is not fundamentally different from the least-squares estimation of equation (3.4), it just allows a robust treatment of correlated data when the covariance matrix is not perfectly known. The transformed variables and their transformed (and now diagonal) covariance matrix yield a modified $\chi^2$ statistic, and if the input data had a multivariate Gaussian distribution we could use this $\chi^2$ value in the standard way to compute confidence intervals on $\beta$ and $\sigma_v$, being careful to reduce the degrees of freedom in the fit by the number of data combinations discarded during PCA. However, even if the multipole estimates $P_2(k)$ and $P_0(k)$ are Gaussian distributed (which they might be, since they involve sums over many wavevectors and many independent windows), their ratio is Gaussian distributed only in the limit that the uncertainty $\Delta P_0$ in the monopole is much smaller than the monopole itself. Thus, although our fitting procedure yields "internal" error estimates from $\chi^2$ values, we obtain our primary assessment of errors on $\beta$ and $\sigma_v$ by applying our method directly to mock 1.2-Jy and QDOT catalogues. This application also allows us to test for any systematic biases in our procedure.

## 4 TESTS ON MOCK CATALOGUES

Using the exponential model described in § 2.1 and the analysis procedure described in § 3, we now examine a set of mock galaxy catalogues. These enable us to define the covariance matrix that that is used in the model fitting and to assess our accuracy in recovering $\beta$.

We construct mock catalogues from the N-body simulations described in § 2.3 by placing the observer at a random location within the simulation cube and selecting particles according to a selection function. We create mock catalogues corresponding to three different surveys:

---

[†] More precisely, since the covariance matrix is symmetric it can be diagonalized by a unitary transformation, $\mathbf{M}' = \mathbf{R}\mathbf{M}\mathbf{R}^\mathbf{T}$, where $\mathbf{M}'$ is a diagonal matrix. The columns of the transformation matrix, $\mathbf{R}$, contain the eigenvectors of $\mathbf{M}$. The linear combinations of the original data points (assumed here to have zero mean) given by $\vec{x}' = \mathbf{R}^T \vec{x}$ are independent because their covariance matrix is diagonal, i.e. $\langle \vec{x}\vec{x}^T \rangle = \mathbf{M}'$. If $\mathbf{M}$ is truly singular, an analogous procedure based on Singular Value Decomposition can be used (cf., Press  et. al. 1992).

 (i) The 1.2-Jy catalogue with the selection function of Fisher et. al. (1995),
 (ii) The QDOT catalogue with the selection function of Lawrence et. al. (1994), and
 (iii) A hypothetical $B_J < 18$ mag all-sky catalogue with the selection function derived by Maddox  et. al. (1990) from the APM survey.

We use periodic replicas of the fundamental simulation cube where necessary, though for the 1.2-Jy and QDOT catalogues the $400h^{-1}$ Mpc simulation cubes are large enough to encompass the survey volume that contributes significantly to our final results.

Catalogues (i) and (ii) enable us to determine the covariance matrix that we should use when analysing the *IRAS* surveys, and we can also use them to estimate errors on the value of $\beta$ determined from these samples. The all-sky $B_J < 18$ mag catalogues (each containing about $1.1 \times 10^6$ galaxies) give some idea of the precision attainable with the next generation of redshift surveys, e.g. the AAT 2dF survey and the Sloan survey. We construct two mock catalogues from each N-body simulation, giving us a total of 8 catalogues for each selection function. Ideally we would use separate simulations for each mock catalogue, but because the noise in our anisotropy measurements arises mainly from the random orientations of real, individual structures — orientations that are different for different "observers" in the simulation volume — the two catalogues per simulation should be effectively independent for our purposes.

We analyse the mock catalogues in exactly the same way as the actual data in the next section. As outlined in § 3, we repeatedly select galaxy subsamples using a spherical window placed randomly within the survey volume. For each set of mock catalogues, the shot noise subtracted power spectrum, $P^S(k,\mu)$, is estimated by direct Fourier transform at the wavelengths corresponding to the fundamental mode and first harmonic of the sampling window. We use seven different sized sampling windows, with fundamental wavelengths $\lambda_f = 140, 120, 100, 80, 60, 40,$ and $20h^{-1}$ Mpc. For each wavelength, the estimates of $P^S(k,\mu)$ from different windows are averaged using the weights defined in § 3.1. The centres of the samples are chosen at random out to the distance at which the weight curve $N_s w_\lambda(d)$ falls to 1% of its peak value (see Figures 3 and 4). There is little point in taking samples at greater depth because they make an insignificant contribution to the weighted average $P^S(k,\mu)$. Two further constraints are placed on the positioning of the samples. First, a minimum distance of $d_{min} \geq R_{\mathrm{sph}}/\tan(25°) = 1.53\lambda_f$ is imposed, so that the angle subtended by each sample is less than 50°. Second, in order to model the excluded regions near the Galactic plane in the actual *IRAS* surveys, windows are rejected if they overlap with $|b| < 5°$ in the mock 1.2-Jy catalogues and $|b| < 10°$ in mock QDOT catalogues. No Galactic cut is imposed in the all-sky $B_J < 18$ catalogues. We use 32,000 samples at each window radius, a number determined by requiring that the $P^S(k,\mu)$ estimates converge.

To estimate $\beta$ we fit monopole and quadrupole terms to each average $P^S(k,\mu)$, correct for the effect of convolution, and fit the exponential 2-parameter model of § 2.1 by the procedure described in § 3.3. Figure 5 shows the average quadrupole-to-monopole ratio derived from the mock cata-logues as a function of wavelength. In each panel, the dots denote the mean values of the ratio from the eight mock catalogues with the indicated selection function; error bars denote the $1-\sigma$ uncertainty in this mean value. The solid curves show the model fit to the mean ratios. For all three selection functions, the model produces a reasonably good fit to the data. On large scales, $\lambda \gtrsim 50 h^{-1}\,\mathrm{Mpc}$, the ratios from the mock QDOT and 1.2-Jy surveys become very noisy. The fitted values of $\beta$ are 0.44 (1.2-Jy), 0.50 (QDOT), and 0.47 ($B_J < 18$). These should be compared with the true value of $\beta = 0.486$ used to construct the simulations.

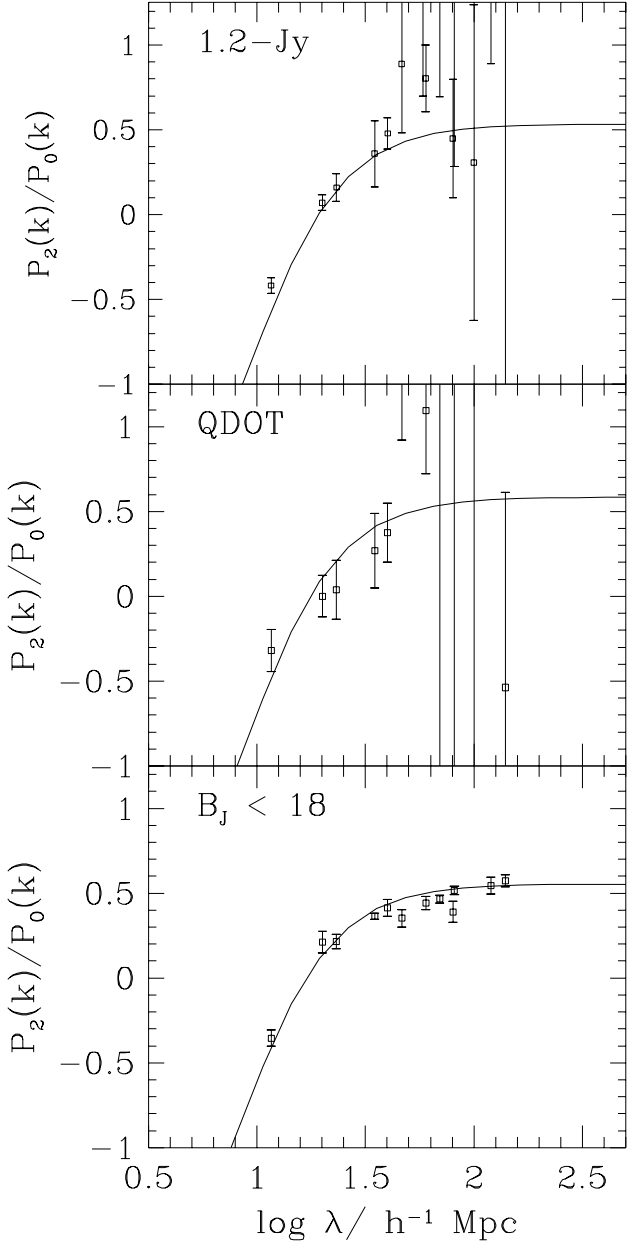

**Figure 5.** The quadrupole to monopole ratios $P_2(k)/P_0(k)$ for the mock catalogues. In each panel we show the average of eight mock catalogues; the error bars denote the $1-\sigma$ uncertainty in the mean. The solid curves are the model fit to the data using the principal component method outlined in § 3.3 . The fitted parameter values are $\beta = 0.45$ and $\sigma_v = 286\,\mathrm{km\,s^{-1}}$, for the mock 1.2-Jy surveys, $\beta = 0.50$ and $\sigma_v = 279\,\mathrm{km\,s^{-1}}$, for the mock QDOT surveys, and $\beta = 0.47$ and $\sigma_v = 261\,\mathrm{km\,s^{-1}}$ for the mock optical surveys with $B_J < 18$.

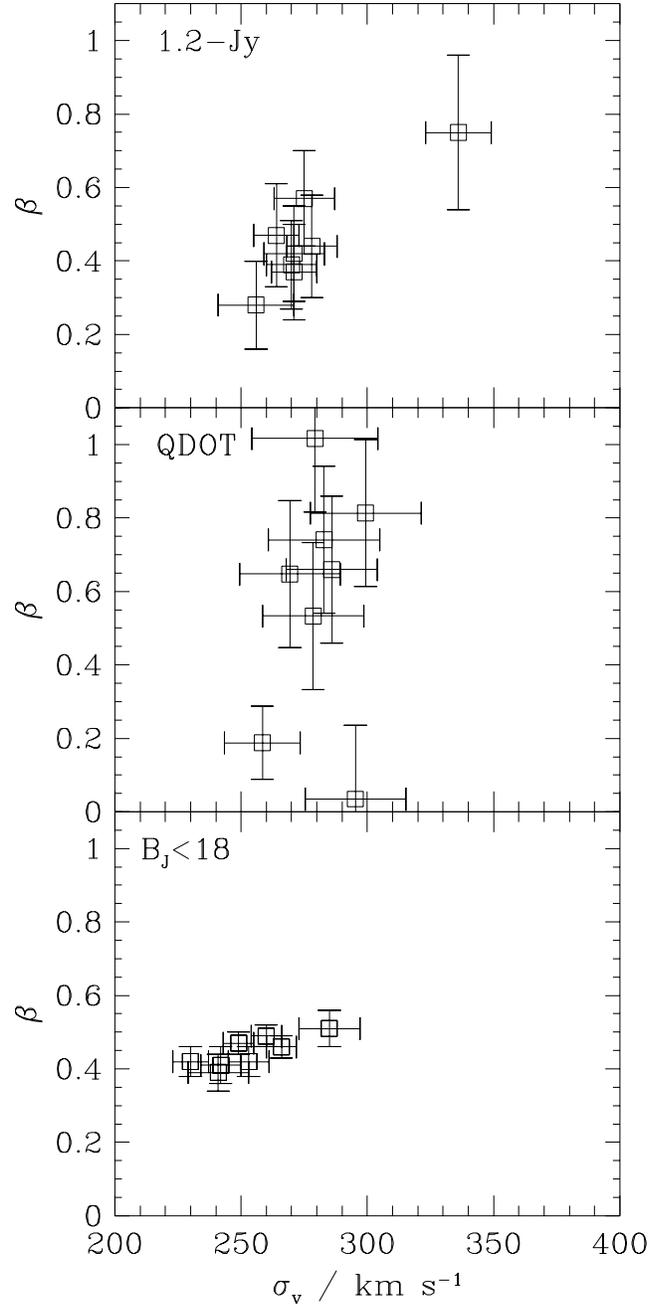

**Figure 6.** The distribution of fitted $\beta$ and $\sigma_v$ parameters for mock catalogues of the a) 1.2-Jy, b) QDOT, and c) $B_J < 18$ surveys. Each point shows the best-fit $\beta$ and $\sigma_v$ values from a single mock catalogue, with the "internal" error bars derived from the $\chi^2$ fitting procedure.

Of course we have only one real 1.2-Jy catalogue and one real QDOT catalogue to analyse, and for individual mock catalogues the typical errors in the quadrupole-to-monopole ratio are larger than the error bars in Figure 5 by a factor of $\sqrt{8-1} \approx 2.6$. Figure 6 shows the distribution of the fitted $\beta$ and $\sigma_v$ values for each of the eight mock catalogues analysed in turn. Each point carries the $1-\sigma$ "internal" error bars derived from the $\chi^2$ contours on the assumption of Gaussian input data (see § 3.3). The mean and dispersion of the $\beta$ estimates are $\beta = 0.46 \pm 0.14$ for the 1.2-Jy catalogues, $\beta = 0.58 \pm 0.32$ for the QDOT catalogues, and $\beta = 0.45 \pm 0.04$ for the $B_J < 18$ catalogues. Estimates of $\beta$ and $\sigma_v$ are correlated, as one might expect, since raising $\beta$ enhances line-of-sight clustering while raising $\sigma_v$ suppresses it.

Figure 6 has a number of encouraging features. First, our method appears to yield estimates of $\beta$ with little if any systematic bias. The mean values obtained from the 1.2-Jy and QDOT mock catalogues differ from the true value $\beta = 0.486$ by less than $0.3\sigma$, where $\sigma$ is the run-to-run dispersion in the $\beta$ estimates. The mean value from the $B_J < 18$ catalogues is about $1\sigma$ below the true value, marginal evidence for a weak systematic effect. Since the points in Figure 2, which come from FFT analysis of the full N-body cubes, lie below the model curve with the true value of $\beta$, we suspect that the weak tendency to underestimate $\beta$ reflects residual non-linear effects on the scales used in the model fit. However, other statistical or geometrical biases could also have an impact. In any event, the mock catalogue results indicate that any systematic errors in the analysis are not important relative to the random errors in surveys the size of 1.2-Jy and QDOT.

The second encouraging feature of Figure 6 is the moderate scatter in $\beta$ values for the 1.2-Jy mock catalogues. The $1-\sigma$ disperson is about 25% of the true $\beta$ value, not negligible by any means, but small enough to make this method competitive with other techniques for estimating $\beta$ (see § 6). The scatter for the QDOT catalogues is larger, mainly because of the larger uncertainties in $P_2(k)/P_0(k)$ at shorter wavelengths. Results for the $B_J < 18$ mock catalogues are impressive, with a run-to-run dispersion of less than 10%. Furthermore, it is clear from Figure 5 that we are not yet taking full advantage of the statistical power in these catalogues; error bars on $P_2(k)/P_0(k)$ are still small at our maximum wavelength $\lambda = 140 h^{-1}\,\mathrm{Mpc}$, and we could improve our $\beta$ determinations by including data at longer wavelengths. Proper investigation of this point will require mock catalogues that better mimic the geometry of planned surveys and that represent long wavelength Fourier modes more accurately than our $400 h^{-1}\,\mathrm{Mpc}$ periodic boxes allow.

A third encouraging feature of Figure 6 is the reasonable agreement between the internal error bars determined from the $\chi^2$ procedure and the external error bars obtained from the run-to-run dispersion. The average internal $1-\sigma$ errors on $\beta$ are 0.14, 0.19, and 0.04 for the 1.2-Jy, QDOT, and $B_J < 18$ catalogues, respectively, while the corresponding external errors are 0.14, 0.32, and 0.04. This concordance suggests that the distribution of the quadrupole-to-monopole ratios is fairly well described by Gaussian statistics.

We can also use mock catalogues to quantify a systematic effect that is neglected in our modelling of the redshift-space distortions. A galaxy's flux, and hence its probability of passing the threshold of a flux-limited sample, depends on its actual distance, not on its redshift *per se*. We should therefore weight each galaxy by the inverse of the selection function evaluated its true position. However, a redshift survey does not provide us with this information, so we must instead evaluate the selection function at the galaxy's redshift-space position. This leads to a distortion in the redshift-space clustering whose sign depends on the gradient of the selection function (see Kaiser 1987). This systematic error is present in the analysis of the mock catalogues presented above, and the fact that we recover the correct value of $\beta$ to an accuracy of 10% suggests that it is not very important. To verify this conclusion, we construct a second set of mock 1.2-Jy catalogues in which we populate the mock survey volumes according to the selection function evaluated at each galaxy's redshift-space position. Our analysis of this second set of surveys is therefore exact. Comparing the $\beta$ values estimated from this second set of mock catalogues with those shown in Figure 6 reveals that they have a very similar distribution, with a mean displaced by $\Delta\beta < 0.05$. Thus this systematic error is small compare to the random errors for surveys the size of the current *IRAS* catalogues. However, the small statistical error for the $B_J < 18$ catalogues shows that it will be important to model this effect when analysing much larger future surveys.

## 5 APPLICATION TO THE 1.2-JY AND QDOT REDSHIFT SURVEYS

We now apply the method we have developed to analyse redshift-space distortions to the two large redshift surveys constructed from the *IRAS* point source catalogue. These are the flux-limited 1.2-Jy survey (Fisher et. al. 1995), which contains 5304 galaxies at Galactic latitudes $|b| > 5°$, and the deeper, sparsely sampled QDOT survey of Lawrence et. al. (1994), which contains 2374 galaxies at $|b| > 10°$. We analyse these data sets in the same way as the mock catalogues described in the previous section. In addition to the Galactic plane region, the 1.2-Jy and QDOT surveys have small excluded areas at high latitudes; we populate these regions randomly with a redshift distribution consistent with the corresponding survey's selection function.

Figure 7 shows the ratio, $P_2(k)/P_0(k)$, of estimated quadrupole and monopole moments as a function of wavelength for both the 1.2-Jy and QDOT surveys. The error bars placed on each estimate come from the mock catalogue analysis described in § 4; they are the error bars of the corresponding panels in Figure 5 expanded by $\sqrt{7}$. The best fit

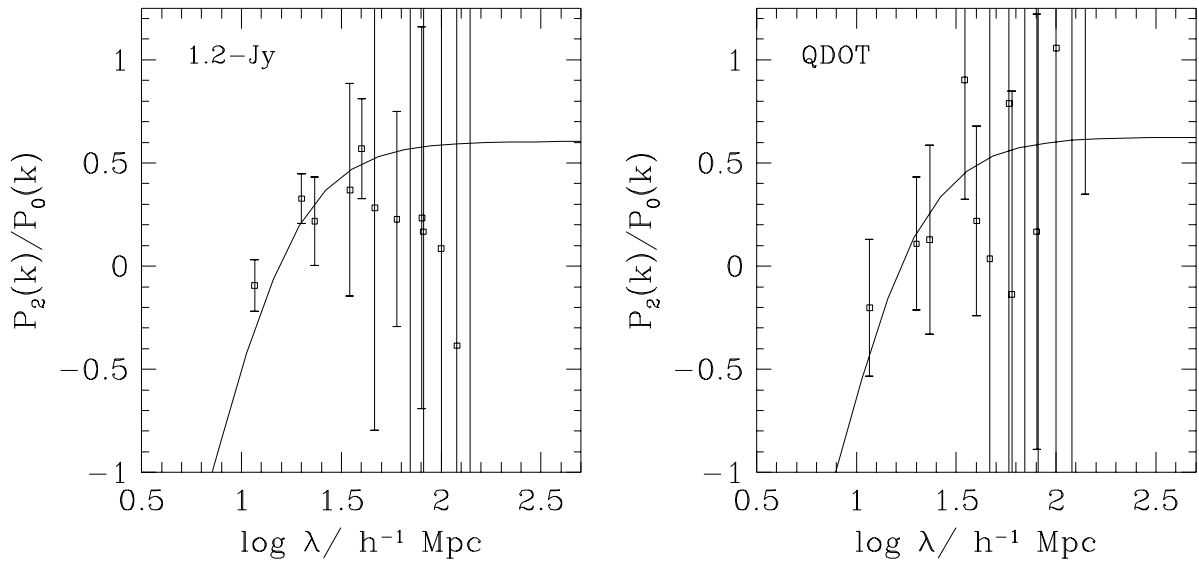

**Figure 7.** The quadrupole-to-monopole ratios, $P_2(k)/P_0(k)$, as a function of wavelength for the actual a) 1.2-Jy, and b) QDOT and catalogues. The error bars placed on these data points are run-to-run dispersion found in the mock catalogues with the same selection function (cf., figure 6) described in section 4. The curves are the model fits to the data made using the full covariance matrix and the method of principal component analysis described in section 3.3. The best fit values of $\beta$ are $\beta = 0.52$ for the 1.2-Jy survey and $\beta = 0.54$ for the QDOT survey.

values of $\beta$ and $\sigma_v$ are

$$1.2 - \text{Jy} : \begin{cases} \beta = 0.52 \pm 0.14 \\ \sigma_v = 249 \pm 25 \text{ kms}^{-1} \end{cases}$$

$$\text{QDOT} : \begin{cases} \beta = 0.54 \pm 0.32 \\ \sigma_v = 276 \pm 13 \text{ kms}^{-1} \end{cases} \quad (5.1)$$

for the 1.2-Jy and QDOT survey respectively. The model fits that give rise to these estimates use the principal component analysis method described in § 3.3, which takes account of correlation between $P_2(k)/P_0(k)$ estimates at different wavelengths. Equation (5.1) quotes $1-\sigma$ errors on $\beta$ and $\sigma_v$ from the run-to-run dispersion of the mock catalogue measurements shown in Figure 6. We list the dispersion parameter $\sigma_v$ mainly for completeness; it bears no simple relation to other conventional measures of small-scale dispersion. The derived values of $\sqrt{2}\sigma_v \sim 350 - 390 \text{ kms}^{-1}$ are in the same ballpark as the pairwise dispersion $\sigma_{pw} = 317^{+40}_{-49} \text{ kms}^{-1}$ estimated by Fisher et. al. (1994b) from the 1.2-Jy redshift-space correlation function. However, as discussed in § 2.3, there is no reason to expect precise agreement between these numbers.

Our PCA fitting procedure (§ 3.3) yields internal error estimates if one assumes that the input data values obey Gaussian statistics. We regard the estimates from the mock catalogues as more reliable than the internal errors, but in fact the two approaches yield similar results: internal $1-\sigma$ errors are $\Delta\beta = 0.15$ (1.2-Jy) and $\Delta\beta = 0.30$ (QDOT) and $\Delta\sigma_v = 15 \text{ km s}^{-1}$ (1.2-Jy) and $\Delta\sigma_v = 25 \text{ km s}^{-1}$ (QDOT). PCA retains seven independent data combinations for the 1.2-Jy survey and six for the QDOT survey, and the two-parameter fits yield $\chi^2$ per degree of freedom of 4.5/5 and 1.8/4, respectively, for the two surveys. The possibility that the quadrupole-to-monopole ratios are not Gaus-sian distributed means that one should not overinterpret these numbers, but they do suggest that our model offers an adequate fit to the data. From Figure 7 it is clear that most of the weight in our estimates comes from wavelengths $10 < \lambda < 40 h^{-1} \text{Mpc}$, and it is the fact that the 1.2-Jy survey yields smaller errors than the QDOT survey in this regime that accounts for its much tighter constraints on $\beta$. Despite the substantial uncertainty in the QDOT estimate, it is reassuring to see consistency between results from the two surveys.

If we assume that $IRAS$ galaxies trace mass, $b = 1$, then equation (5.1) implies

$$\Omega = 0.34^{+0.16}_{-0.14} \text{ } (1.2 - \text{Jy}), \quad \Omega = 0.36^{+0.42}_{-0.28} \text{ (QDOT)}. \quad (5.2)$$

If, on the other hand, we assume $\Omega = 1$, then we find a bias factor for $IRAS$ galaxies of

$$b = 1.92^{+0.71}_{-0.40} \text{ } (1.2 - \text{Jy}), \quad b = 1.85^{+2.70}_{-0.69} \text{ (QDOT)}. \quad (5.3)$$

Optical galaxies are more strongly clustered than $IRAS$ galaxies, and the corresponding bias factor for optical galaxies would be higher by a factor of about 1.4 (see, e.g., Fisher et. al. 1994a).

## 6 DISCUSSION

In this paper, we have extended the formalism introduced in Paper I to include the effect of small-scale, non-linear velocity dispersion, using a simple, 2-parameter model for redshift-space distortion, and we have presented a practical statistical method for treating flux-limited surveys. We have applied the method to the $IRAS$ 1.2-Jy and QDOT surveys, from which we estimate $\beta = 0.52 \pm 0.14$ and $\beta = 0.54 \pm 0.32$ respectively. The errors quoted here are the 1-$\sigma$ uncertainty deduced from ensembles of mock catalogues.

**Table 1.** Constraints on $\beta$ deduced from the $IRAS$ galaxies surveys.

| Method | $\beta$ | Survey | Reference |
| --- | --- | --- | --- |
| Anisotropy of $P^s(k,\mu)$ | $0.52 \pm 0.15$ | 1.2-Jy | This work. |
| Anisotropy of $P^s(k,\mu)$ | $0.54 \pm 0.3$ | QDOT | This work. |
| Anisotropy of $\xi(\mathbf{s})$ | $0.69^{+0.28}_{-0.24}$ | 2.0-Jy | Hamilton (1993) |
| Anisotropy of $\xi(\mathbf{s})$ | $0.45^{+0.27}_{-0.18}$ | 1.2-Jy | Fisher et. al. (1994b) |
| $\xi(\mathbf{s})$ vs. $w(\theta)$ | $1.0 \pm 0.2$ | QDOT | Peacock & Dodds (1994) |
| $\xi(\mathbf{s})$ vs. $w(\theta)$ | $0.84 \pm 0.45$ | QDOT | Fry & Gaztañaga (1994) |
| Spherical Harmonics | $0.94 \pm 0.17$ | 1.2-Jy | Fisher et. al. (1994c) |
| Spherical Harmonics | $1.1 \pm 0.3$ | 1.2-Jy | Heavens & Taylor (1994) |
| Dipole | $0.82 \pm 0.15$ | QDOT | Rowan-Robinson et. al. (1991) |
| Dipole | $0.55^{+0.2}_{-0.12}$ | 1.2-Jy | Strauss et. al. (1992b) |
| Differential Dipole | 0.6 | 1.2-Jy | Nusser & Davis (1994) |
| Peculiar velocities | $0.86 \pm 0.14$ | QDOT | Kaiser et. al. (1991) |
| $\delta$ vs. $\nabla \cdot \mathbf{v}$ | $1.28^{+0.38}_{-0.3}$ | 2-Jy | Dekel et. al. (1993) |

Methods based on the two-point correlation function and on decomposition of the redshift-space density field into spherical harmonics offer alternative approaches to measuring $\Omega$ and $b$ from the anisotropy of clustering in redshift space. Each of these approaches admits a number of variants. For example, one can estimate $\beta$ from the anisotropy of the redshift-space correlation function, $\xi(r_p, r_\pi)$ using linear theory (Hamilton 1993) or models that extend to the non-linear regime (Fisher et. al. 1994b). Alternatively, one can estimate $\beta$ from the enhancement of the angle-averaged redshift-space correlation function over the real-space correlation function inferred by deprojecting the angular correlation function $w(\theta)$. There are even variants to measuring the anisotropy of the redshift-space spectrum. Instead of Fourier transforming windowed subsamples of the data as we do here, Hamilton (1995) estimates $P^s(k,\mu)$ by a direct sum over galaxy pairs, imposing the small-angle constraint on the pairs individually. Pairwise estimation may make more efficient use of the redshift data than our window technique, but this point has yet to be examined in detail.

Spherical harmonics (Fisher, Scharf, & Lahav 1994c; Heavens & Taylor 1994) do not require a small-angle constraint because they use truly radial eigenfunctions. They therefore provide the best way to utilize the largest scales probed by a survey. For the $IRAS$ surveys this is a major advantage, because only modes close to the effective depth of the survey are accurately in the linear regime. Other approaches to redshift-space anisotropy require non-linear modelling, as we have done in this paper. Spherical harmonics are also well suited to the nearly full-sky coverage of the $IRAS$ surveys. The method requires modification for samples that cover only a fraction of the sky, but it is still possible to perform decomposition with radial eigenfunctions (e.g., Vogeley 1994), thereby maintaining the attractive feature of this approach.

One can also estimate $\beta$ by comparing the gravitational accelerations predicted from the galaxy density field to the dipole motion of the Local Group or to peculiar velocities inferred from redshift-independent distance measurements. Table 6 lists various estimates of $\beta$ for $IRAS$ galaxies. An expanded table of $\beta$ and $\Omega$ estimates and a thorough discussion of methods that incorporate peculiar velocities appear in the excellent review by Strauss & Willick (1995). It is clear from Table 6 that a consensus on the value of $\beta$ has not yet been reached. The various estimates of $\beta$ are roughly consistent with each other given the quoted errors, but there is considerable scatter in the values of $\beta$ estimated from the same survey. This suggests that for some or all of the methods, either random errors have been underestimated or residual systematic errors remain. The spherical harmonic analyses of the $IRAS$ catalogues yield $\beta$ values that are significantly higher than those obtained from anisotropy of the power spectrum or the correlation function. However, the values of $\beta$ from the spherical harmonic analyses are sensitive to the normalization of the $IRAS$ galaxy power spectrum; if this normalization is included as a free parameter in the method, then lower values of $\beta$ can be obtained (Fisher et. al. 1994c).

Estimates of $\beta$ from peculiar velocity data should improve in the near future, as larger velocity data sets become available and more sophisticated analysis methods provide a better handle on systematic effects. However, it is difficult to push peculiar velocity samples to large depths because the error in a galaxy's estimated velocity scales in proportion to its distance. Redshift surveys, on the other hand, are very much a growth industry, and it seems likely that in 3−5 years the best constraints on $\beta$ will in fact come from measurements of clustering anisotropy. Our tests on all-sky, $B_J < 18$ mock catalogues, which contain $\sim 1.1 \times 10^6$ galaxies apiece, suggest the power that these techniques have when applied to very large redshift surveys. The third panel of Figure 6 shows that the expected uncertainty in $\beta$ for such a sample is only $\Delta\beta \sim 0.04$, and this error could probably be reduced further by using longer wavelength data. The mean value of $\beta$ derived from these mock catalogues appears to be biased low, probably because of residual non-linearity not accounted for by our two-parameter model. Full exploitation of future redshift samples will require more sophisticated theoretical models, and by detecting other non-linearities one can hope to break the degeneracy between $\Omega$ and $b$ (see discussion in Paper I).

It is not clear what approach to redshift-space

anisotropy — power spectrum, correlation function, or modified spherical harmonics — will be the most powerful for redshift samples like the AAT 2dF and Sloan Digital Sky Surveys. These surveys will have excellent statistics on scales that are close to the linear regime but still much smaller than survey volume, and it is these scales that one will want to exploit. To take full advantage of these data, further investigation is required on both statistical estimation and dynamical modelling, but current results suggest that the reward will be worth the effort.

## ACKNOWLEDGMENTS

We thank Changbom Park for the use of his $N$-body code and John Huchra, Michael Strauss, Marc Davis and Amos Yahil for allowing us the use of the 1.2 Jy survey prior to publication. SC acknowledges the support of a PPARC advanced fellowship. KBF acknowledges the support of the Ambrose Monell Foundation. DHW acknowledges the support from the W.M. Keck Foundation and NSF grant PHY92-45317.